\documentclass[prb,twocolumn,aps,showpacs,10pt]{revtex4}
\usepackage{graphicx}
\usepackage{bm}
\usepackage{amssymb} 
\usepackage{color}

\begin{document}

\title{Resonant finite-size impurities in graphene, unitary limit and Friedel oscillations}

\author{V. V. Mkhitaryan and E. G. Mishchenko}

\affiliation{Department of Physics and Astronomy, University of Utah, Salt Lake
City, UT 84112, USA}

\begin{abstract}
Unitary limit for model point scatterers in graphene is known to
reveal low-energy resonances. The same limit could be achieved
from hybridization of band electrons with the localized impurity
level positioned in the vicinity of the Fermi level. The finite
size defects represent an easier realization of the effective
unitary limit, occurring when the Fermi wavelength induced by the
potential becomes of the order of the size of the defect. We
calculate the induced electron density and find two signatures of
a strong impurity, independent of its specific realization. The
dependence of the impurity-induced electron density on the
distance changes near resonances from $\propto r^{-3}$ to $\propto
r^{-2}$. The total number of induced particles at the resonance is
equal to one per degree of spin and valley degeneracy. The effects
of doping on the induced density are found.
\end{abstract}

\maketitle

\section{Introduction}

In conventional three-dimensional electron systems probe charges
are screened exponentially with the distance\cite{PN}. In
degenerate metals at distances exceeding the screening radius the
non-monotonic power-law tail develops, $\propto
\cos(2k_{\scriptscriptstyle F}r)/r^3$, that originates from
electron backscattering with the change of momentum equal to
$2k_{\scriptscriptstyle F}$, twice the Fermi momentum
\cite{Friedel}. Correspondingly, in a conventional two-dimensional
electron gas\cite{LK} the amplitude of these Friedel oscillations
decays as only the second power of the distance $\propto r^{-2}$.
In systems with the Berry phase and  non-trivial chiral spectrum
Friedel oscillations could still decay faster, $\propto r^{-3}$,
if the states with momenta ${\bf k}$ and $-{\bf k}$ are orthogonal
and backscattering is suppressed\cite{CR}. In particular, this
happens in case of a doped graphene \cite{CN}, when the Fermi
level is shifted away from the Dirac points,
$k_{\scriptscriptstyle F}\ne 0$, with pseudospin-diagonal impurity
potential. For a short-range impurity with the potential $V({\bf
r})=U\delta({\bf r})$ the resulting induced electron density is
\cite{CF, BV} (we use units with $\hbar =1$)
\begin{equation}\label{FOgraph}
n({\bf r}) = \frac{U}{2\pi^2 vr^3}\cos(
2k_{\scriptscriptstyle F}r), ~~~k_{\scriptscriptstyle F}r \gg 1,
\end{equation}
here $v$ is the velocity of Dirac electrons. Within the linear
response approximation the derivation of this result is
straightforward with the help of the density-density correlation
function $\Pi_\omega ({\bf r}-{\bf r'})$ taken in the static
$\omega=0$ limit: $n=\Pi_0 \otimes V$. In case of a short-range
potential the latter formula gives simply $n({\bf r}) =U \Pi_0
({\bf r})$, and eventually leads to Eq.~(\ref{FOgraph}).

The calculation outlined above can be extended easily to the case
of intrinsic graphene, $k_{\scriptscriptstyle F}= 0$, or,
equivalently, to short distances, $r\ll k_{\scriptscriptstyle
F}^{-1}$. Interestingly, the functional form (\ref{FOgraph}) is
recovered again, as can be verified within the same linear
response approach; the only change being in the numerical factor
$\pi/4$:
\begin{equation}\label{FOgraph1}
n({\bf r}) =  \frac{U}{8\pi vr^3},~~~ k_{\scriptscriptstyle
F}r \ll 1.
\end{equation}
The non-integrable singularity at $r\to 0$ makes the total
induced electron density $\int n({\bf r})d^2r $ diverge as
a power law. This  indicates the failure of the first Born approximation
when $U/vr \sim 1$. Fortunately, the delta-function potential allows for an
exact non-perturbative solution via the $T$-matrix in terms of the
electron Green's function, $T(E) = U/ [1-U \sum_{\bf
p}G(E,{\bf p})]$, which yields,
\begin{equation}
\label{t-matrix-delta} T(E)=\frac{U}{1+\frac{U}{2\pi v^2}E
\bigl[\ln(v/a|E|)+i\pi/2\bigr]}.
\end{equation}
The ultraviolet divergence is cut-off at short distances of the
order of the lattice spacing $a$.

In the "unitary limit" of strong interaction, $U \to \infty$, the
resonant form of the $T$-matrix (\ref{t-matrix-delta}) clearly should
lead to a significant modification of the dependence of the
induced density on distance. It is not difficult to obtain a rough
estimate for the effect: since typical energies are $|E|\sim v/r$,
we immediately conclude that in the logarithmic approximation,
$n(r) \sim {1}/[r^2 \ln{(r/a)}]$. Surprisingly, as we demonstrate
in the present paper this crude guess is incorrect and the
induced density in the unitary limit is in fact {\it suppressed}
according to ($U>0$)
\begin{equation}
\label{resonant_density} n(r) = \frac{2}{\pi r^2
\ln{(U/va)}}.
\end{equation}
The behavior described by Eq.~(\ref{resonant_density}) cannot be
properly accounted for if the imaginary part of $T(E)$ is
neglected. It turns out that despite the imaginary part in the
denominator of Eq.~(\ref{t-matrix-delta}) being small compared
with the logarithmically large real part, the two parts lead to
the contributions that largely cancel each other.

The suppression of the local density (\ref{resonant_density}) at
the unitary limit does not mean that the {\it total} density
vanishes too. Quite to the contrary, as the range of distances,
$r\ll U/v$, where Eq. (\ref{resonant_density}) is applicable,
increases with increasing $U$ the total induced density tends to a
limit $\int n({\bf r})d^2r \to 4$, i.e. {\it one electron} per
spin/sublattice.

The resonant $1/E$-behavior of the scattering amplitude in graphene
has significant implications. In particular, the unitary limit leads
to singular corrections to the low-energy density of states
\cite{OGM, PGCN, SPG}. The effective Casimir-like coupling between
two point impurities (adatoms) is predicted to become long-range in
the unitary limit \cite{SAL} with the potential energy $\propto
[r\ln{(r/a)}]^{-1}$. However, achieving the unitary limit by the
strength of the potential alone might be difficult. To illustrate
this for atomic impurities let us consider the simplest microscopic
model of a point impurity that yields the expression
(\ref{t-matrix-delta}), namely, the tight-binding approximation for
the honeycomb lattice where a single carbon atom is substituted with
an impurity atom \cite{BK, Basko}. In that case the effective
strength of the delta-function is simply $U \sim \langle V_i \rangle
a^2$, where $a=1.4\AA$ is the interatomic distance and $\langle V_i
\rangle$ is the expectation value of the impurity's potential energy
(calculated with the help of the unperturbed orbital). We therefore
estimate that for the unitary limit to be effectively achieved at
distances $r$ the strength of the impurity should
exceed\begin{equation} \label{estimate_U} \langle V_i \rangle \gg t
~\frac{r/a}{\ln{(r/a)}},
\end{equation}
with $t=v/a\approx 3$ eV being the hopping energy. From the
expression (\ref{estimate_U}) it follows that already at distances
$r\sim 1-2$ nm the impurity strength $\langle V_i \rangle$ needs
to be of the order of tens of eV, which is clearly impractical.

This restriction could be lifted if an impurity has a localized
level with low energy $\varepsilon_0 \ll t$ leading to a resonant
enhancement of scattering \cite{WKL}. The limit of strong
impurities could also be realized in vacancies \cite{PGC06}. Yet
another possibility is a finite-size scatterer \cite{HG,TOG}.
Indeed, consider an impurity potential of finite radius $\rho$.
Treating potential energy perturbatively is justified as long as
the characteristic potential energy $V_0$ is small compared with
the typical kinetic energy, $v/\rho$. The latter estimate uses
that the typical electron momenta inside the potential in {\it
intrinsic} graphene are determined by the width of the potential,
$k\sim 1/\rho$. The Born approximation,
Eqs.~(\ref{FOgraph})-(\ref{FOgraph1}), then holds true for small
effective dimensionless coupling constants, $g=V_0\rho/v \ll 1$.
The non-perturbative regime is reached when $g\sim 1$, which is a
{\it much less stringent} (by a factor $r/a$) condition than the
above discussed condition for a delta-function impurity. This is
also evident from the correspondence $\pi V_0\rho^2 \to U$
expected to exist between the potentials at $\rho \to 0$.

In the present paper we address the response of Dirac electrons in
graphene to strong impurities. Since the case of a finite-size
scatterer reveals the richest behavior and is free from low-distance
singularities we are going to address it first. Still, the results
obtained will be applicable to other strong impurities with proper
modifications.  In Section II we begin with analyzing the square
potential problem, $V({\bf r}) = V_0\Theta (\rho-r)$, and finding
the induced density in case of intrinsic graphene.  The unitary
limit is resonantly achieved whenever the coupling constant
coincides with a zero of a Bessel function, $J_0(g)=0$, where the
induced density changes its dependence on the distance to $\propto
1/r^2$. Physically, resonances are related to the number of
wavelengths induced by the potential $V_0$ that fit inside it. Such
a possibility is absent for point-like scatterers where the strength
of the potential is the only variable parameter that has to be sent
to infinity in order to achieve the unitary limit. In Section III we
address the effects that occur due to finite Fermi momentum in doped
graphene, including Friedel oscillations.

\section{Impurity-induced electron density in intrinsic graphene}

Solutions of the Dirac equation in a centrally symmetric potential
$V(r)$ have been considered extensively before since
Refs.~\onlinecite{HG,Novikov,Basko}, but for the sake of
convenience we present them here again. The Dirac equation
\begin{equation}\label{Deq}
-iv \left(\begin{array}{cc}
0&\partial_x-i\partial_y\\ \partial_x+i\partial_y& 0
\end{array}\right)\psi=[E-V(r)]\psi,
\end{equation}
determines the two-component wavefunction is a linear combination of
partial waves with the angular momentum quantum number $m=\pm 1/2, \pm
3/2,..$
\begin{equation}\label{partwave}
\psi_m=\frac{e^{im\phi}}{\sqrt{2\pi}}\left(\begin{array}{c}
\varphi_1e^{-i\phi/2}\\ \varphi_2e^{i\phi/2}
\end{array}\right).
\end{equation}
Here we introduced the polar coordinates: $x+iy =re^{i\phi}$.

\begin{figure}[t]
\centerline{\includegraphics[width=90mm,angle=0,clip]{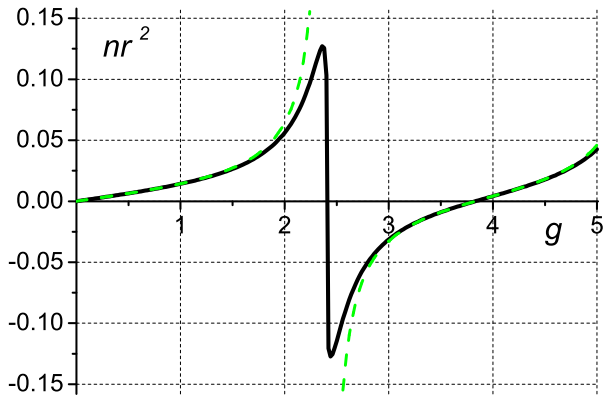}}
\caption{(Color online) Dependence of the induced density
(rescaled by $r^2$) on the dimensionless impurity strength,
$g=V_0\rho/v$, plotted from Eq. (\ref{densviaK}) for $r/\rho=10$.
The approximation which is valid away from the resonance, Eq.
(\ref{tailviag}), is shown  for comparison with the green dashed
line. The most notable feature is the reversal of the sign of the
induced density from positive (electrons) to negative (holes) as
$g$ passes through the point where $J_0(g)=0$. Near  the
``anti-resonance'' point $g\approx 3.8$, determined by $J_1(g)=0$
the impurity becomes ``invisible'' as the induced density is
strongly suppressed by a small factor $\rho/r$.}\label{mainres}
\end{figure}

We first focus on the repulsive potential $V_0>0$. As the integral
over all filled states (negative energies $E$) has to be taken
eventually, we parameterize $E=-vk$ with positive $k$. Then the
Schr\"odinger equation (\ref{Deq}) gives two coupled equations for
$r<\rho$,
\begin{eqnarray}
&&\partial_r\varphi_1-\frac{m-1/2}r\varphi_1=
-i\left(k+\frac{V_0}{v}\right)\varphi_2,\nonumber \\
&&\partial_r\varphi_2+\frac{m+1/2}r\varphi_2= -i\left(k+
\frac{V_0}{v} \right)\varphi_1.\label{down}
\end{eqnarray}
Similarly, for $r>\rho$ the wave functions obey the same equations
as Eqs.~(\ref{down}) but without $V_0$. Both inside and outside of
the potential the wave functions satisfying the condition of
regularity at $r=0$ are given in terms of Bessel functions,
\begin{equation}\label{inwf}
\left(\begin{array}{c}
\varphi_1\\
 \varphi_2
\end{array}\right)=A\left(\begin{array}{c}
J_{m-1/2}\left(kr+gr/\rho\right)\\
-iJ_{m+1/2}\left(kr+gr/\rho\right)
\end{array}\right)
\end{equation}
for $r<\rho$, and
\begin{equation}\label{outwf}
B\!\left(\begin{array}{c}
\text{sgn}(k) J_{m-1/2}(|k|r)\\
-i J_{m+1/2}(|k|r)
\end{array}\right) +C\!\left(\begin{array}{c}
\text{sgn}(k)Y_{m-1/2}(|k|r)\\
-i Y_{m+1/2}(|k|r)
\end{array}\right)
\end{equation}
for $r>\rho$. The coefficients $A$, $B$, and $C$ are found from the
conditions of continuity at $r=\rho$. The normalized \cite{norm}
wave functions for $m=1/2$ in the outer region, $r>\rho$, are then
found to be
\begin{eqnarray}\label{normWF}
&&\varphi_1=\text{sgn}(k)\frac{\sqrt{\pi
|k|}}{\sqrt{\beta^2+\gamma^2}} \Bigl[
\beta J_0(|k|r)+\gamma Y_0(|k|r)\Bigr], \nonumber\\
&&\varphi_2=-i \frac{\sqrt{\pi |k|}}{\sqrt{\beta^2+\gamma^2}}
\Bigl[ \beta J_1(|k|r)+\gamma Y_1(|k|r)\Bigr],
\end{eqnarray}
where the following coefficients are defined ($k>0$),
\begin{eqnarray}\label{bcviaa}
&&\beta= J_1\left(k\rho+g\right)Y_0(k\rho)-
J_0\left(k\rho+g\right)Y_1(k\rho), \nonumber\\
&&\gamma=J_0\left(k\rho+g\right)J_1(k\rho)-
J_1\left(k\rho+g\right)J_0(k\rho).
\end{eqnarray}
The electron density $n=\frac{2}{\pi^2} \int dk (|\varphi_1|^2
+|\varphi_2|^2)$ takes into account the contributions from the two
leading channels with $m=\pm 1/2$ as well as spin/valley degeneracy.
Subtracting the equilibrium ($g=0$) density we find,
\begin{eqnarray}\label{dens1}
n(r)&=&
\frac2{\pi}\sum_{i=0,1}\int\limits_0^\infty\frac{dk\,k}{\beta^2+\gamma^2}
\Bigl\{ 2\beta\gamma\ J_i(kr)Y_i(kr)
\nonumber\\
&&+\gamma^2\left[Y_i^2(kr)-J_i^2(kr)\right] \Bigr\},
\end{eqnarray}
The latter equation can be written more compactly if we use the
Hankel function, $H^{(1)}(z)=J(z)+iY(z)$, to obtain
\begin{equation}\label{densviaH}
n(r)=
\frac2{\pi}\text{Im}\sum_{i=0,1}\int\limits_0^\infty\frac{dk\,k\gamma}{\beta+i
\gamma} \left[{H_i^{(1)}}(kr)\right]^2.
\end{equation}
We are now going to analyze this expression for different values
of the impurity strength $g$. The oscillating behavior of the
Hankel functions ensures that the integral in Eq.~(\ref{densviaH})
converges at $k\sim 1/r$. Since we are interested in the behavior
at long distances $r \gg \rho$ we can write for generic values of
$g$ (though still $\gg \rho/r$),
\begin{equation}
\label{gammaapprox} \gamma\approx
-J_1(g).
\end{equation}
The latter approximation is valid unless $g$ is very close to a
zero of the Bessel function $J_1(g)$ (see below). Similarly, by
keeping the singular terms in the small-argument expansion of the
Bessel functions we can write for the second coefficient,
\begin{eqnarray}\label{betaapprox}
\beta&\approx& J_1(g) Y_0(k\rho) -J_0(g)Y_1(k\rho)\nonumber\\
&\approx& \frac 2\pi\left[J_1(g)\ln(k\rho)
+J_0(g)\frac1{k\rho}\right].
\end{eqnarray}
Substituting approximate Eqs.~(\ref{gammaapprox}-\ref{betaapprox})
into the exact formula (\ref{densviaH}) we arrive at the induced
density,
\begin{equation}\label{densviaHexp}
n(r)= \frac1{r^2}\text{Im}\sum_{i=0,1}\int\limits_0^\infty\frac{dz
\,z \left[{H_i^{(1)}}(z)\right]^2} {\displaystyle \ln
\left(\frac{r}{\rho z}\right)-\frac {J_0(g)}{J_1(g)}\frac{r}{\rho
z}+i\frac\pi2}.
\end{equation}
Here we introduced $z=kr$. The integral is most easily calculated
by deforming the integration contour so that it follows the
positive half of the imaginary axis. The Hankel functions have a
branching point at $z=0$ and a cut that extends along the negative
real axis. With such a choice of the branch cut the logarithm in
the denominator takes the value $\ln(r/y\rho)-i\pi/2$ on the
positive part of the imaginary axis ($z=iy$). The Hankel functions
become Macdonald functions via
$\left[{H_i^{(1)}}(iy)\right]^2=\frac{4}{\pi^2}(-1)^{i+1}\left[K_i(y)\right]^2$.
We therefore arrive at
\begin{equation}\label{densviaK}
n(r)=-\frac4{\pi^2r\rho}\frac{J_0(g)}{J_1(g)} \int\limits_0^\infty
\frac{dy~ [K_0^2(y)-K_1^2(y)]}{\displaystyle
\ln^2\left(\frac{r}{\rho y}\right)+ \left(
\frac{J_0(g)}{J_1(g)}\frac{r}{\rho y} \right)^2}.
\end{equation}
Due to the exponentially suppressed MacDonald functions, this
integral is dominated by $y\lesssim 1$. Then, unless the
combination $J_0(g)/J_1(g)$ is small, the logarithm in the
denominator can be safely neglected. Utilizing the numerical value
of the integral, $\int_0^\infty dy y^2
[K_0^2(y)-K_1^2(y)]=-\pi^2/16$, leads to the following expression,
\begin{equation}\label{tailviag}
n(r)= \frac{J_1(g)}{J_0(g)} \frac \rho {4r^3}.
\end{equation}
This equation determines a nontrivial $g$-dependence, represented
in Fig.~\ref{mainres}. For weak impurities, $g\ll 1$,
Eq.~(\ref{tailviag}) reproduces the first Born approximation for
the delta-function (\ref{FOgraph1}), if one takes into account the
obvious correspondence for the strength of the delta-function,
$\pi v g \rho \to U$. For strong impurities $g>1$ the induced
density is not positively defined and can become {\it negative},
which happens for example when $2.4<g<3.8$, see
Fig.~\ref{mainres}.

When the impurity strength obeys the equation $J_1(g)=0$, the
induced density is strongly suppressed, e.g. for $g=3.8$. In such
a case it is no longer possible to use the approximation
(\ref{gammaapprox}). From Eq.~(\ref{bcviaa}) we now get $\gamma
\approx k\rho (J_0/2-J_1')$. The presence of an extra power of $k$
ensures much faster decay of the electron density \begin{equation}
n(r)\propto \frac{\rho^2}{r^{4}}.
\end{equation}

So far we have discussed the contributions from the lowest order
$s$-wave scattering, $m=\pm 1/2$. A few words are now in order about
higher $m$ channels. The equations (\ref{normWF}-\ref{bcviaa}) are
fully applicable there as long as the order of the Bessel functions
is adjusted: $0\to m-1/2$, $1 \to m+1/2$ (for positive $m$). Instead
of Eqs.~(\ref{gammaapprox}-\ref{betaapprox}) we now have $\gamma
\propto (k\rho)^{m-1/2}$, $\beta \propto 1/(k\rho)^{m+1/2}$. From
Eq.~(\ref{dens1}) we observe that each extra order of $m$ brings
additional factor $(k\rho)^2$. Upon taking the $k$-integral we
conclude that already the $p$-wave scattering contribution is
suppressed by a small factor $(\rho/r)^2$ and could be safely
neglected. It is also known that for $|m|\geq 3/2$ zero energy bound
states exist\cite{BTB} when $J_{|m|-1/2}(V_0\rho)=0$. However, as
the electron density for these states decays $\propto r^{-2|m|-1}$
their contribution to $n(r)$ is negligible.

\subsection{The case of resonant scattering}

\begin{figure}[t]
\centerline{\includegraphics[width=90mm,angle=0,clip]{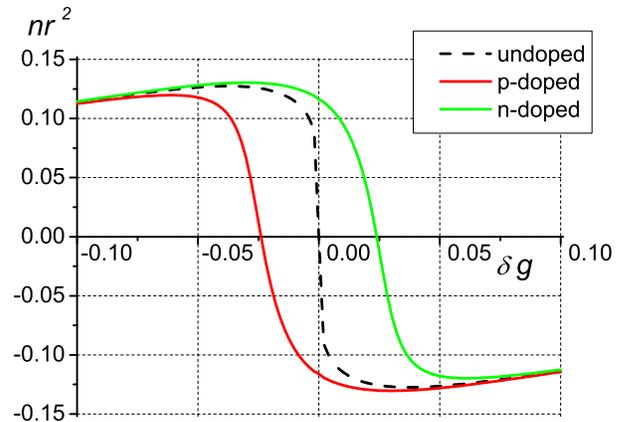}}
\caption{(Color online) Sensitivity of the resonant behavior to
small levels of doping, $k_Fr=0.05$, and $r/\rho=10$. Induced
density $n(r)r^2$ is plotted near $g=g_c+\delta g$. The black
dashed line represents the behavior of induced density in undoped
graphene. In case of $p$-doping the resonance is shifted by
$\Delta g_c=k_{\scriptscriptstyle F}\rho\ln{(k_{\scriptscriptstyle
F}\rho)}<0$. Similarly, in case of an $n$-doped system the changes
are reversed (for the discussion of the $n$-doped graphene see
Section III.C).} \label{nearres}
\end{figure}

With increasing $g$ the system passes through a set of resonances
\cite{Ablyazov} determined by the condition $J_0(g_c)=0$, the
first of which occurs at $g_c=2.4$, Fig.~\ref{mainres}. The
physical origin of these resonances is quite clear and could be
identified with the number of potential-induced Fermi wavelengths
$v/V_0$ that fit inside the radius of the well $\rho$. In the
vicinity of the resonance $J_0(g)$ in Eq.~(\ref{densviaK}) could be
expanded as
\begin{equation}
J_0(g)\approx -J_1(g_c)\delta g, ~~~~ g=g_c+\delta g.
\end{equation}
Contrary to the calculations of the preceding section, the logarithm
in the denominator of Eq. (\ref{densviaK}) can not be neglected
anymore. In fact the main contribution to the integral comes from
small values of $y$. The  numerator mostly comes from
$K_1^2(y)\approx y^{-2}$, yielding,
\begin{equation}\label{newregdens1}
n(r)=-\frac{4\delta g }{\pi^2r\rho}\int\limits_0^1
\frac{dy}{\displaystyle y^2\ln^2\left(\frac{r}{\rho
y}\right)+\left(\delta g\frac{r}{\rho} \right)^2}.
\end{equation}
In agreement with the assumption just made the arguments  relevant
in the integral (\ref{newregdens1}) are small, $y\sim \delta
gr/(\rho |\ln{\delta g}|) \ll 1$. With the logarithmic accuracy
(neglecting double logarithm) we obtain,
\begin{equation}\label{critdens}
n(r)=\frac{2}{\pi r^2} \frac{\text{sgn} (\delta g)}{ \ln{|\delta
g|}}, ~~~ |\delta g| \ll \frac{\rho}{r}\ln{\left(\frac{r}{\rho}
\right)}.
\end{equation}
This relation describes the replacement of the growth of the
density when a resonance is approached, see Eq.~(\ref{tailviag}),
with the ultimate logarithmic suppression in the immediate
vicinity of it. The dependence of the density on distance changes
from $r^{-3}$ away from the resonance to
 $r^{-2}$ near it. The maximum of the density occurs at impurity
strength $|\delta g| \sim ({\rho}/{r})\ln{({r}/{\rho})}$ where the
two expressions (\ref{critdens}) and (\ref{tailviag}) match, as
could be expected. The behavior of the induced density in the
vicinity of resonance is presented in Fig.~\ref{nearres}.

It is now interesting to compute the total induced electron density.
Such a calculation can be most rigorously carried out by integrating
the general expression (\ref{densviaK}): it is convenient to first
integrate over $r$ (from $\rho$ to $\infty$) and then over $y$. The
same results, however, could much faster be found straight from
Eqs.~(\ref{tailviag}) and (\ref{critdens}). Far from the resonance,
$g\ll 1$, integrating Eq.~(\ref{tailviag}) we obtain,
$N_{tot}=\frac{\pi}{4}g$. As expected, in this perturbative regime
the total disturbance of the system is small. Near the resonance,
$\delta g \ll 1$, upon integrating Eq.~(\ref{critdens}) from $\rho$
to $\rho/|\delta g|$ we arrive at
\begin{equation}
N_{tot} =-4~\text{sgn}(\delta g).
\end{equation}
Thus, the total number of particles induced by the resonant impurity
is equal to {\it one per Dirac cone} (considering spin and valley
degeneracy) and changes sign right at the resonance. This conclusion
is in agreement with the Friedel sum rule\cite{Friedel52}.

\subsection{Delta-function potential}

We now discuss a connection between the finite size well and the
point impurity $V({\bf r})=U\delta({\bf r})$. As already evident
from Eq.~(\ref{tailviag}) there is no rigorous limit of
progressively narrower and higher potential, $g=V \rho/v \propto
1/\rho \to \infty$, as $\rho$ tends to zero. Alternatively, this
could also be seen from the fact that the delta-function potential
does not allow a dimensionless coupling parameter independent of
the distance $r$ of the kind that $g$ is for a finite-size well.
Yet, there is a connection between the two models.

First we note that the combination of $\beta$ and $\gamma$ in the
integrand of Eq.~(\ref{densviaH}) is proportional to the
$T$-matrix. Indeed, by taking the asymptotics of the wave function
(\ref{normWF}) the scattering phase shift could be expressed as,
$\tan{\delta} =-\gamma/\beta$. The relation between the phase
$\delta$ and scattering amplitude in graphene is very similar to
the standard quantum mechanical expression\cite{LLIII} and was
written in Ref.~\onlinecite{Novikov}. In our notations,
\begin{equation}
\label{t-matrix-phase} T(k)=-\frac{4v}{k}
\frac{\gamma}{\beta+i\gamma}.
\end{equation}
Comparing now this general expression with
Eq.~(\ref{t-matrix-delta}) we can identify the phase shift for a
point impurity,
\begin{equation}
\label{betaovergamma}
\cot\delta=-\frac{\beta}{\gamma} =\frac{4v}{kU} -\frac{2}{\pi}
\ln\left(\frac{1}{ka}\right).
\end{equation}
From Eqs.~(\ref{gammaapprox}) and (\ref{betaapprox}) we can now
extract the correspondence between the two problems,
\begin{equation}
\label{correspondence} U ~\longleftrightarrow ~2\pi v\rho
\frac{J_1(g)}{J_0(g)}.
\end{equation}
The combination in the right-hand side (without $v$) could be
identified with the scattering length of the circular potential
well\cite{TOG}.

We observe that the correspondence already encountered away from
the resonances, cf.~Eqs.~(\ref{FOgraph1}) and (\ref{tailviag}),
holds in fact everywhere. The finite-size resonances discussed in
the previous section thus represent realizations of the unitary
limit. Replacing now  $\delta g\to -2\pi v\rho/U$ in Eq.
(\ref{critdens}) we arrive at Eq.~(\ref{resonant_density}) up to a
replacement of $\rho$ with $a$ under the logarithm; this
distinction is beyond logarithmic accuracy anyway.

It should be emphasized that the imaginary part in the scattering
matrix Eq.~(\ref{t-matrix-delta}) {\it cannot be neglected}. This
can be seen following the transformation from
Eq.~(\ref{densviaHexp}) to Eq.~(\ref{densviaK}), where it is
important that the imaginary part is cancelled by the phase coming
from the logarithm upon the rotation to the imaginary axis.
Without $i\pi/2$ in the denominator of Eq.~(\ref{densviaHexp}),
the density would have been finite in the unitary limit instead of
the suppression described by Eq.~(\ref{resonant_density}).

Note that the correspondence (\ref{correspondence}) is different
from what one would expect to be the true delta-functional limit: $U
\leftrightarrow \pi v g\rho$ with $\rho \to 0,~g\to \infty$.
Interestingly, the two models coincide with each other only in the
case of a {\it shallow} well, $g\ll 1$. This situation is unique for
the linear spectrum of graphene. Indeed, in the case of a
conventional parabolic spectrum, $\epsilon =p^2/2m^*$ the
dimensionless phase shift $\delta = \delta (V_0\rho^2m^*)$ could
only be a function of the combination $V_0 \rho^2$ that becomes
$U/\pi$ in the limit of a deep and narrow well. This could be
qualitatively understood by recalling that the parameter $Um^*$
gives the number of bound states in a well of depth $V_0=U/\pi
\rho^2$, which does not depend on $V_0$ or $\rho$ {\it separately}.
In graphene, to the contrary, the phase shift is a function of
$g=V_0\rho/v$.


A closely related to the delta-function potential is the case of
point-like Anderson impurity with a low-energy state $\varepsilon_0$
hybridized with graphene ban electrons via $H_{int}=u(d^\dagger
\psi(0)+{c.c.})$, the problem considered in Ref.~\onlinecite{WKL}.
When $\varepsilon_0\approx 0$ the $T$-matrix has the resonant form
(\ref{t-matrix-delta}) with the strength of the potential replaced
by $U \to u^2/\varepsilon_0$. Similar substitution in
Eq.~~(\ref{resonant_density}) yields the distribution of the induced
density. Note that the density is positive as long as
$\varepsilon_0>0$ (electrons are expelled from the vicinity of the
impurity) and negative when $\varepsilon_0<0$. Hydrogen adatoms are
known to have resonances very close to the Dirac
point\cite{WKL2009}, $\varepsilon_0\approx 30~\text{meV}$.


\section{Extrinsic graphene, $k_{\scriptscriptstyle F}\ne
0$}

Let us now consider the case of gated or doped graphene with a
nonzero Fermi-momentum. We will analyze two cases separately. When
the distance to the impurity exceeds the Fermi wavelength,
$k_{\scriptscriptstyle F}r\gg 1$, the conventional Friedel
oscillations develop whose specific behavior depends strongly on
the impurity strength. The opposite case of a weakly gated
graphene, $k_{\scriptscriptstyle F}r\ll 1$, is much more
spectacular. Because of the resonant behavior described by
Eqs.~(\ref{tailviag}) and (\ref{resonant_density}) the sensitivity
of the induced density to small $k_{\scriptscriptstyle F}$ turns
out to be very strong.

\subsection{Weakly $p$-doped or gated graphene, $k_Fr\ll 1$}

The case of a $p$-doped graphene is particularly straightforward
as it is sufficient to simply replace  the lower limit in the
integral in Eq.~(\ref{densviaHexp}) with $k_{\scriptscriptstyle
F}r$. The integral is then written as
$\int_0^\infty-\int_0^{k_{\scriptscriptstyle F}r}$, with the first
one yielding the same expression as before,
Eq.~(\ref{resonant_density}). The second integral brings the
correction,
\begin{equation}
\label{correction} \Delta n(r)=-\frac{2}{\pi
r}\int\limits_0^{k_Fr} \frac{zdz}{\displaystyle
\Bigl[z\ln\left(\frac{r}{\rho z}\right)+\delta g\frac{r}{\rho}
\Bigr]^2+\frac{\pi^2}{4}z^2}.
\end{equation}
From the form of this integral it is clear that the correction is
highly {\it asymmetric}: it quickly decays with increasing $\delta
g$ when the latter is positive. For negative $\delta g<0$ the
correction (\ref{correction}) is much stronger as the integrand
has a pole near the real axis. The pole is significant as long as
$2k_{\scriptscriptstyle F}\rho\ln{(k_{\scriptscriptstyle F}\rho)}<
\delta g <0$. Outside of this region the correction again
decreases quickly. This is illustrated in Fig.~\ref{nearres}: the
plot of the total density $n_{tot}(r)=n(r)+\Delta n(r)$ is
effectively "shifted" to the left of $g_c$. The first terms in the
Taylor expansion in powers of $\delta g$ could be easily
extracted:
\begin{equation}
\label{densitytotal} n_{tot}(r) =-\frac{2}{\pi
r^2\ln{(k_{\scriptscriptstyle F}\rho)}}
\left(1+\frac{(\pi-2)\delta g}{k_{\scriptscriptstyle F}\rho
\ln^2{(k_{\scriptscriptstyle F}\rho)}} \right).
\end{equation}
The shift of the position of the resonance due to doping, $\Delta
g_c$, with great precision  is given by the condition,
$\beta(k_{\scriptscriptstyle F})=0$, yielding
\begin{equation}
\Delta g_c= -k_{\scriptscriptstyle F}\rho
\ln\left(\frac1{k_{\scriptscriptstyle F}\rho}\right),
\end{equation}
see also discussion in the next Section. Another consequence of
Eq.~(\ref{densitytotal}) is that the singularity in the derivative
of the undoped resonant density is regularized by finite
$k_{\scriptscriptstyle F}$.

Let us emphasize the signature feature of the resonant behavior
depicted in Fig.~\ref{nearres}, namely the {\it reversal} of the
sign of the density via small doping occurring in the range
$\Delta g_c<\delta g<0$.

\subsubsection{Delta-function potential}

\begin{figure}[t]
\centerline{\includegraphics[width=90mm,angle=0,clip]{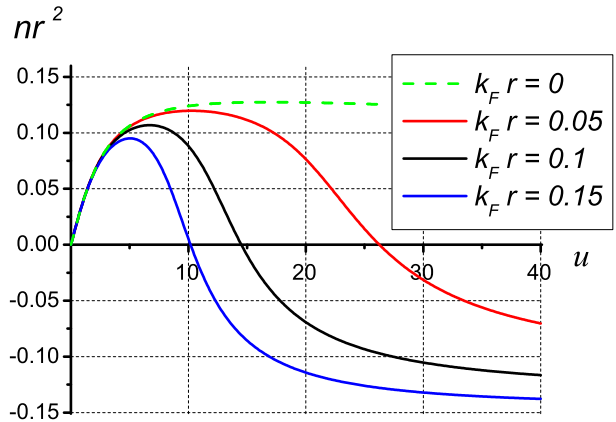}}
\caption{ (Color online) The emergence of a resonance for a
point-like impurity described by the delta-function potential
$V(r)=U\delta({\bf r})$ for weak $p$-doping levels (for $U>0$)
versus the dimensionless impurity strength $u=U/vr$, plotted from
Eq.~(\ref{density_total}), for $r/a=10$. The green dashed line
illustrates the undoped case: the logarithmic decay
(\ref{resonant_density}) is very weak on the scale shown. Finite
$k_{\scriptscriptstyle F}$ makes the density change sign at $U_0$
given by Eq.~(\ref{U_0}). The same resonance is revealed via
$n$-doping for an attractive potential of the impurity, $U<0$.}
\label{deltapot}
\end{figure}

While the effects of weak doping lead to significant qualitative
modification to the dependence of induced density on the distance
for finite-size impurities, the corresponding changes for point
defects are even more drastic. Note that the induced density
in the latter case is {\it always} of the same sign as $U$ in case
of intrinsic graphene. In other words the delta-function potential
roughly maps on the interval $0<g<2.4$, below the first resonance
of the finite-size well, which corresponds to the $U\to \infty$
unitary limit, cf.~Eq.~(\ref{correspondence}). Doping with small
amounts of holes (for $U>0$) or electrons (for $U<0$), on the other
hand, {\it pushes} the resonance to a {\it finite} impurity
strength $U_0$, so that for $U>U_0$ the induced particle density
{\it changes} sign, much like in the case of a finite-size well.
This strong modification of the induced density is captured in
Fig.~\ref{deltapot}.

Although we explained below Eq.~(\ref{correspondence}) that
obtaining expressions for $U\delta({\bf r})$ is rather simple, let
us present here for reference the total induced density for the
latter case,
\begin{eqnarray}
\label{density_total}
n_{tot}(r)&=&-\frac{8}{\pi r^2 u}\int\limits_0^\infty dy \frac{
K_0^2(y)-K_1^2(y)}{\displaystyle \ln^2\left(\frac{r}{a y}\right)+
\left(
\frac{2\pi}{u y} \right)^2}\nonumber\\
&&-\frac{2}{\pi r^2} \int\limits_0^{k_Fr} \frac{zdz}{\displaystyle
\Bigl[z\ln\left(\frac{r}{a z}\right)-\frac{2\pi}{u}
\Bigr]^2+\frac{\pi^2}{4}z^2},~~~~
\end{eqnarray}
where by $u=U/vr$ we denoted the dimensionless parameter of the
problem. Fig.~\ref{deltapot} illustrates the dependence of the
induced density on $u$ for different values of doping levels
$k_Fr$. The density changes sign at the point where
\begin{equation}
\label{U_0} U_0= -\frac{2\pi v}{k_{\scriptscriptstyle
F}\ln(k_{\scriptscriptstyle F} a)}.
\end{equation}
At $U=U_0$ the scattering phase shift at the Fermi level
$\delta_{\scriptscriptstyle F}$, determined by
Eq.~(\ref{t-matrix-phase}) undergoes a jump from $\pi/2$ to
$-\pi/2$.

\subsection{Strongly $p$-doped case, $k_Fr\gg 1$}

With further increase in the doping level the vicinity of the
Fermi surface begins to dominate. Eq.~(\ref{densviaH}) with the
lower limit replaced by $k_Fr$ is a convenient starting point.
Using the asymptotic expression
$\sum_{i=0,1}[{H_i^{(1)}}(z)]^2=-(\frac{2}{\pi z^2})\exp{(2iz)}$
and noting again that the ratio $\gamma/\beta=-\tan{\delta}$ is
related to the phase shift $\delta$ we obtain,
\begin{equation}
n(r)= \frac{4}{\pi^2 r^2} \text{Im} \int\limits_{k_Fr}^\infty
\frac{dz}{z}~\sin\delta~ e^{2iz+i\delta}.
\end{equation}
Because of the strongly oscillating behavior of the integrand only
the vicinity of the lower limit contributes to the density. After
simple integration we arrive at,
\begin{equation}\label{densviadelta}
n(r) = 2\sin\delta_{\scriptscriptstyle
F}\frac{\cos\bigl(2k_{\scriptscriptstyle
F}r+\delta_{\scriptscriptstyle F}\bigr)} {\pi^2
k_{\scriptscriptstyle F}r^3},
\end{equation}
with  the scattering phase shift taken at the Fermi surface,
$\delta_{\scriptscriptstyle F}=-\tan^{-1}[
\gamma(k_{\scriptscriptstyle F})/\beta(k_{\scriptscriptstyle
F})]$. The formula (\ref{densviadelta}) describes Friedel
oscillations both in the perturbative limit where
$\delta_{\scriptscriptstyle F}\ll 1$ and Eq.~(\ref{FOgraph}) is
recovered, and the resonant (unitary) regime, when
$\delta_{\scriptscriptstyle F}$ is close to $\pm \pi/2$ where the
amplitude of the oscillations no longer depends on the impurity
strength. Note that in the vicinity of the resonance the density
(\ref{densviadelta}) reverses its sign, similar to the intrinsic
case.

\subsection{$n$-doped graphene}

Now we discuss $n$-doped graphene with ground state filled up to a
positive Fermi energy, $E_{\scriptscriptstyle F}>0$. According to
our notation, $k=-E/v$, in the $n$-doped case the range of
positive energies with $k_{\scriptscriptstyle F} <k<0$ appears in
addition to the filled lower Dirac cone. For negative $k$ the
eigenfunctions Eq.~(\ref{normWF}) are still applicable with
coefficients now assuming the form
\begin{eqnarray}\label{kFbcviaa}
&&{\beta}= J_1\left(k\rho+g\right)Y_0(|k|\rho)+
J_0\left(k\rho+g\right)Y_1(|k|\rho), \nonumber\\
&&{\gamma}=-J_0\left(k\rho+g\right)J_1(|k|\rho)-
J_1\left(k\rho+g\right)J_0(|k|\rho),\qquad
\end{eqnarray}
implying that Eq.~(\ref{gammaapprox}) holds for $\gamma$ with both
signs of $k$, whereas Eq.~(\ref{betaapprox}) captures correct
$\beta$ for all values of $k$ if one changes $k$ to $|k|$ in the
argument of logarithm. Taking this into account, the induced
density is given by Eq. (\ref{dens1}) or (\ref{densviaH}),
provided that the lower limit of the integral is extended to
$k_{\scriptscriptstyle F}$ and $k$ is changed to $|k|$ in the rest
of the integrand. We write it as
\begin{equation}\label{kFdens1}
n(r)=\frac2{\pi}\text{Im}
\sum_{i=0,1}\left\{\int\limits_{-\infty}^\infty
-\int\limits_{-\infty}^{-|k_{\scriptscriptstyle
F}|}\right\}\frac{dk\,|k|\gamma}{\beta+i \gamma}
\left[{H_i^{(1)}}(|k|r)\right]^2.
\end{equation}
It is straightforward to check that the first integral here is
zero. Indeed, as a combination of integrals over $(-\infty,\,0)$
and $(0,\,\infty)$, it amounts to the sum of two integrals, one of
which is the same as in Eq.~(\ref{densviaHexp}), while the other
one is given by Eq.~(\ref{densviaHexp}) with the opposite sign of
$J_0(g)/J_1(g)$ in the denominator. According to Eq.
(\ref{densviaK}) such a combination is zero.

We conclude that the induced density is governed by the second
integral of Eq.~(\ref{kFdens1}), leading to
relation~(\ref{densviadelta}) with the opposite sign. Therefore,
the general relation for induced density (Friedel oscillations) in
strongly doped graphene is
\begin{equation}\label{generalFO}
n(r) = -\frac{2v\sin\delta_{\scriptscriptstyle
F}}{E_{\scriptscriptstyle
F}}\frac{\cos\bigl(2|k_{\scriptscriptstyle
F}|r+\delta_{\scriptscriptstyle F}\bigr)} {\pi^2 r^3}.
\end{equation}
The phase shift at the Fermi surface is determined by
(cf.~Eq.~\ref{betaovergamma})
\begin{equation}
\label{phaseF} \cot{\delta_{\scriptscriptstyle F}}=-\frac {2v}{\pi
E_{\scriptscriptstyle F}\rho}\frac{J_0(g)}{J_1(g)} -\frac{2}{\pi}
\ln\left(\frac{v}{|E_{\scriptscriptstyle F}|\rho}\right).
\end{equation}
The expression (\ref{generalFO}) generalizes Eq.~(\ref{FOgraph})
and has the latter as the limiting case at weak couplings. The
long-behavior is always $\propto r^{-3}$, even at resonance, as
opposed to the intrinsic graphene, Eq.~(\ref{resonant_density}).
Still, as one passes through a resonance, the phase shift jumps by
$\pi$ and the sign of the induced density reverses, much like in
Fig.~\ref{mainres}.

\section{Summary and Conclusions}

In this paper we considered the two non-perturbative models for
impurities in graphene: a substitution atom described by
$V({\bf r})=U\delta({\bf r})$, and a finite-size impurity (molecule or
nanoparticle) with $V({\bf r})=V_0\Theta(\rho-r)$. The first model cannot
be derived from the second one via a standard limiting procedure in the effective
low-energy Dirac fermion description, although there is a simple
correspondence between the two. As an
illustration of the scattering problem we calculated the
impurity-induced density both for intrinsic and extrinsic (doped)
graphene.

The case of a finite-size potential reveals a set of resonances
that correspond to the zeros of the Bessel function, $J_0(g)=0$,
for the dimensionless impurity strength, $g=V_0\rho/v$, see
Fig.~\ref{mainres}. Near a resonance (which is an effective
realization of the unitary limit) the $r^{-3}$ decay of the
induced density is replaced with a slower $r^{-2}$-dependence. The
resonances are interspaced with ``anti-resonances'' that occur
when $J_1(g)=0$. At the latter points the induced density becomes
strongly suppressed with the long-distance behavior $\propto
r^{-4}$. Low dopings, $k_Fr\ll 1$, modify the vicinity of
resonances leading to the shift of the resonant coupling $g_c$ and
removal of the logarithmic singularity, Fig.~\ref{nearres}. The
rest of the $n(g,r)$-dependence is unaffected by low
$k_{\scriptscriptstyle F}$, including positions of the
anti-resonances. At stronger doping levels, $k_Fr \gg 1$, the
induced density follows the usual Friedel
$\cos{(2k_Fr)}/r^3$-dependence known from the first Born
approximation, with the resonant behavior entering via the phase
shifts. In particular, near the resonances the induced density
reverses sign. The Friedel oscillations remain strongly suppressed
near anti-resonances.

The point impurity model $V=U\delta({\bf r})$ does not have
anti-resonances. In the intrinsic case the only true resonance
occurs at $U = \infty$ though the maximum of the induced density
is found near $U\sim 10vr$. With the doping the resonance is
shifted to the finite values $U_0$ given by Eq.~(\ref{U_0}), see
Fig.~\ref{deltapot}.

Our findings indicate that the unitary limit of strong impurities
could be realized with realistic potentials $V_0$. In particular,
for a nanometer-size impurity $\rho\approx 1$ nm the first resonance
occurs when $V_0=2.4 v\hbar /\rho\approx 1.5$ eV. Correspondingly,
for an extended $\rho\approx 10$ nm defect the potential would have
to be only $V_0 \approx 150$ meV.

\acknowledgements Useful discussions with M.~Raikh, O.~Starykh, and
P.G.~Silvestrov are gratefully acknowledged. The work was supported
by the Department of Energy, Office of Basic Energy Sciences, Grant
No.~DE-FG02-06ER46313.


\begin{references}
\bibitem{norm} We normalize the wavefunctions on the "$k/2\pi$ scale",
$ \int\! dr\,r\bigl[\varphi^\ast_{1 k}(r)\varphi_{1 k^\prime}(r)
+\varphi^\ast_{2 k}(r)\varphi_{2
k^\prime}(r)\bigr]=2\pi\delta(k-k^\prime).$
\end{references}
\end{document}